# Stability of glassy hierarchical networks

**M Zamani[1], L Camargo-Forero[2] and T Vicsek[1,3]**

[1]*Department of Biological Physics, Eötvös University, Pázmány P. stny 1/A, 1117 Budapest, Hungary*

[2] *Universitat Politècnica de Catalunya, Barcelona TECH, Esteve Terradas, 5, 08860, Castelldefels, Catalunya, Spain.*

[3] *MTA-ELTE Statistical and Biological Physics Research Group, Pázmány Péter s. 1/A, 1117 Budapest*

Email: vicsek@hal.elte.hu

**Abstract:**

The structure of interactions in most of animals and human societies can be best represented by complex hierarchical networks. In order to maintain close to optimal functioning both stability and adaptability are necessary. Here we investigate the stability of hierarchical networks that emerge from the simulations of an organization-type having an efficiency function reminiscent of the Hamiltonian of spin-glasses. Using this quantitative approach we find a number of expected (from everyday observations) and highly non-trivial results for the obtained locally optimal networks, including such as: i) stability increases with growing efficiency and level of hierarchy, ii) the same perturbation results in a larger change for more efficient states, iii) networks with a lower level of hierarchy become more efficient after perturbation, iv) due to the huge number of possible optimal states only a small fraction of them exhibits resilience and, finally, v) "attacks" targeting the nodes selectively (regarding their position in the hierarchy) can result in paradoxical outcomes.

## 1. Introduction

Stability is one of the most essential features of complex systems ranging from ecological [1-2] to social [3,4], communication [5-7] and economic networks [8] or even multi-robot systems acting as a single collective intelligent system based on the ideas and features of a High Performance Computing Cluster [9]. Stability of a system can be investigated from several perspectives including the perhaps two most essential ones: resistance and resilience. From the resistance point of view, the main question is how resistant a system is against external perturbations. In this regard, networks that persist for longer time in the presence of perturbations are considered more stable or, alternatively, they are also more stable (resistant) if higher magnitude of perturbation is needed to deviate a metastable system from its stationary, locally optimal state. Resilience refers to how quickly a system recovers from disturbance and returns to its equilibrium or stationary state. In both of these approaches, the change of some appropriately chosen variables could be used as a tool for measuring the level of stability.

In Ref. [10] we introduced a model in order to interpret the apparently glassy behaviour of hierarchical organizations and their corresponding network of interactions. Here, glassy behaviour means that according to observations, a given organization (or, in general, most of the complex systems) can maintain several or many metastable states depending on their initial structure and the perturbations they are subject to. The model [10] leads to a complex behaviour of the efficiency function associated with the performance of networked organizations resembling the phenomena displayed by the so called spin glass model [11-13]. Within the above approach, organizations have many local optimal states which are close to each other and, in addition, maximizing the efficiency function leads to hierarchical structure in the networks of the interactions between individual units. Here we address the question of central importance: how stable is the network structure against perturbations? What is the relation between efficiency and stability and how stability can be related to the structure of the network? Are the hierarchical structures more stable than the less hierarchical ones? For answering these questions, we first need to define stability. In standard physical systems stability is defined using second derivative of potential energy [14]: when it is larger than zero (and the first derivative equals to zero) it means that the potential energy is at a local minimum and small perturbation returns the system to its stable state. How can we define such a parameter for complex networks?

Stability of complex networks has been defined in several ways. It has been studied extensively by considering random and targeted nodes and links removal [15-17] based in part on percolation theory. In these approaches, network connectivity is a crucial criterion for measuring stability [18], networks are considered stable if their connectivity is unaffected by the removal of high number of nodes and/or links. Although connectivity is an important feature of networks, their detailed structure and further global properties play fundamental roles in the interaction of individuals in social, robotic and economic systems [19], thus, the latter properties represent significant parameters when measuring stability. Different theoretical models have been developed to understand the formation of social and economic networks and, at the same time, their efficiency and stability have also been analysed [21-22]. Change of network variables and parameters (such as efficiency) due to changing environment or any external disturbance are typically considered as stability measurement criteria. In recent work by Geo et al, the resilience of multi-dimensional networked systems was measured by reducing them to one-dimension [23]. In their approach, it was assumed that networks are in their steady state (fixed points), and because of changing environment they may lose their resilience by sudden transition to other undesired fixed points. Node, weight and link removal are externally imposed perturbations to the system which is an undirected network.

In this paper, the stability of locally optimal states of directed complex networks is examined by adding two kinds of perturbations (noise) to the system. While after optimization the structure of the network freezes in one of its locally optimal states, the effect of noise relocates links or removes nodes in the system. Change in efficiency and global reaching centrality (level of hierarchy) [24] are studied between local and noisy states and are compared for different values of noise. Network resistance against perturbation is investigated by measuring the number of steps that should be taken before disturbing efficient state. Network resilience is considered by looking at the ability of the system to return to its local optimal state after turning off the noise.

## 2. Perturbing networks

In any complex network a general function (describing the total/global state of the system) can be defined. In our previous communication [10], we developed an efficiency function for a typical organization/system which is constructed from interacting individual units having a variety of abilities $a_i$ (level of the potential contribution of a unit to the performance of the whole system). This function reflects the facts that the contribution resulted by the "collaboration" of two units is proportional to their multiplied abilities and can be both positive and negative:

$$E_{eff} = 1/N \sum_{ij} J_{ij} a_i a_j \qquad (1)$$

Where $N$ is the number of nodes. Directed edges between individuals have signs corresponding to their harmonic ($J_{ij} = +1$) or antagonistic ($J_{ij} = -1$) relation. $J_{ij} = 0$ if the two nodes are not connected. The direction of the edges is related to the sign of the expression $a_i - a_j$ (is pointed in the majority of case from the unit with a higher to a unit from a lower ability). Eq. (1) has a similar structure as a spin-glass Hamiltonian. However, there is a notable "twist" in the present interpretation: while in the case of the standard spin-glass framework it is the spins which are varied for finding configurations with small free energy, in Eq. (1) the $J_{ij}$-s (i.e., the network configurations) are tuned to maximize the efficiency, $E_{eff}$, while the $a_i$-s are constant.

The model has 3 parameters, the probability of antagonistic interactions ($q$) and the direction of an edge pointing against the larger ability node ($p$) and, in addition, a constraint for the maximum of the incoming plus the outgoing edges ($in + out$). The results we present are for systems of $N$ nodes, $p = q = 0.2$ and $in + out = 10$. Before the optimization starts a full graph of $N$ nodes with given $J_{ij}$-s and edge directions is generated. Then a subgraph (within the full graph) of $M = 3N$ randomly chosen edges is created. In most of the cases this subgraph has a number of nodes equal to $N$. The efficiency function is maximized in order to find local optimal states of the networks using Monte Carlo simulation. The resulting network efficiencies and their distribution corresponding exhibit a glassy behaviour meaning that the optimization converges to many states. Maximizing the efficiency function leads to complex directed networks with hierarchical features. The distribution of local maxima of efficiencies and their corresponding Global Reaching Centrality ($GRC$) or the level of hierarchy values indicate that optimal states fall into two categories with high and low $GRC$.

Global Reaching Centrality ($GRC$) is defined by following equation [24].

$$GRC = \frac{\sum_{i \in V}[C_R^{max} - C_R(i)]}{N-1}, \qquad (2)$$

where $N$ is the number of nodes in the network, $C_{R(i)}$ is the local reaching centrality of the node $i$ that is described as the number of nodes which can be reached from node $i$ through the directed edges of the network. $C_R^{max}(i)$ is the maximum of $C_{R(i)}$ and the summation is over all nodes in the graph $V$. The question pose is: to what degree is an optimal state reached by maximizing Eq. (1) is stable, from resistance point of view against external perturbation and is their ability to return to their optimal state after turning off the noise (resilience).

In a mechanical system, after a long time particles tend to stay in an equilibrium state and resist against any disturbance from outside. For a highly stable state, higher external force is required to perturb the system permanently out from its equilibrium or a well pronounced metastable state.

A somewhat modified version of this concept is used in this paper for defining the level of stability of complex networks generated by Ref. [10]. Therefore, for a given magnitude of the external perturbations, the number of steps needed to deviate from the local optimal state and the efficiency difference caused by external perturbation or noise can be considered as quantities for checking stability. External perturbation can be a noise in a local optimal state of the system, after optimal states are achieved through Monte Carlo simulation by randomly relocating the position of the edges for temperatures close to zero. In each Monte Carlo step, the efficiency is calculated. If the efficiency is higher than the previous step it is accepted and if it is lower, it is accepted by Boltzmann probability ($\exp\left(-\frac{\Delta E_{eff}}{T}\right)$). To reach the saturated highly efficient state, temperature ($T$) in Boltzmann probability should be close to zero. After reaching the optimal states where efficiency saturates, we increase temperature to implement the noise which increases the Boltzmann probability. The noise is kept on until efficiency changes. Then the difference between efficiencies and $GRC$-s in the two states (optimal and noisy states) as well as the number of steps taken to see the first change in efficiency are calculated. Random relocation of edges in local optimal states is another way for perturbing the system, it is shown that the high value of $T$ or noise has the same effect as of the edge relocation.

Moreover, we investigate the effect of removing nodes (attacks) on the efficiency and $GRC$. An attack is defined as deleting $Q$ number of nodes from the network in its local optimal states, with $Q = [1,2,3, ... , N_l]$, where $N_l$ is the maximum number of nodes removed. For each of the $Q$ attacks, we measure efficiency and $GRC$ after the attack. The effect of external perturbation by adding noise in local optimal states (temperature increase in Boltzmann probability in Monte Carlo simulation) and attack to the system by targeted node removal are depicted schematically in Figure 1.

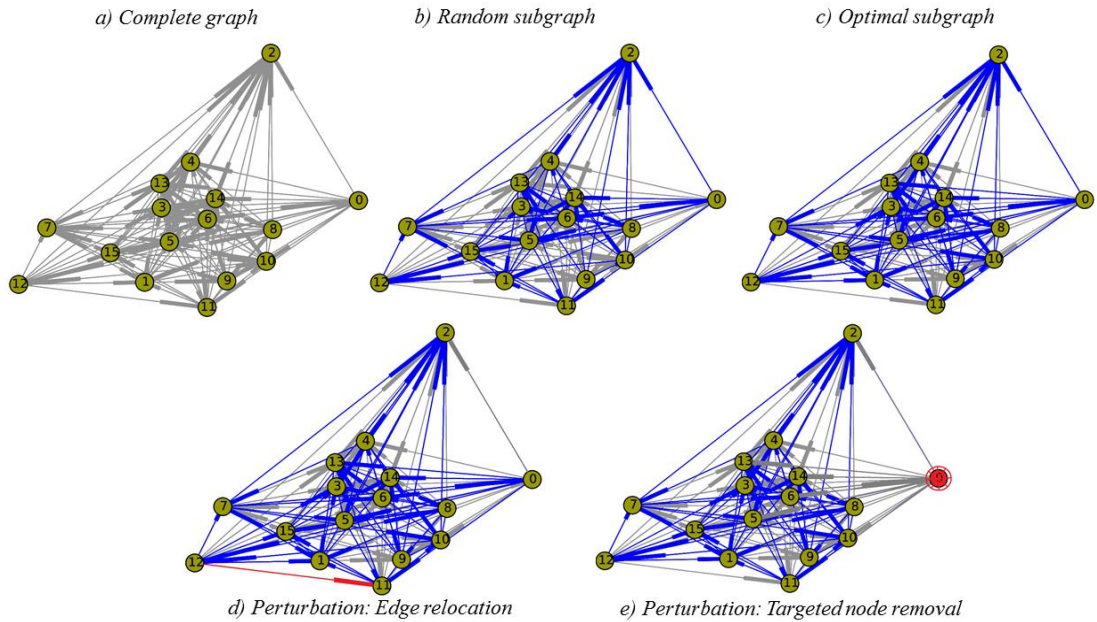

**Figure 1. Illustration of the processes carried out during the simulations.** An initial complete graph with 16 nodes and a random initial subgraph inside it (*a*, *b*). Local optimal state is reached by edge relocation (resulting in a new subgraph) with simultaneous maximizing of its corresponding efficiency. The network structure in one of the local optimal states is shown in (*c*). The effect of perturbation in local optimal state by turning on the noise (by increasing the temperature in the Monte Carlo simulation) is shown in (*d*). The directed edge from node 0 to 2 is removed and the other edge (red) is added from node 12 to 11. Targeted node removal is displayed in (e). Node 0 and its corresponding link is removed from optimal subgraph in (e).

## 3. Results

We perform the following computational experiment to understand the relation between efficiency and stability. We start with a random graph of $M$ edges with a certain efficiency. Then we follow the procedure in Ref. [10] to maximize efficiency as defined in Eq. (1). We consider $u = 100$ different optimal states chosen in a random fashion. For each optimal state, we turn on the noise and Monte Carlo simulation continues until system abandons the optimal state in favour of an unstable state. The number of steps taken in Monte Carlo until the first change in efficiency observed is saved. We perform this measurement $w = 100$ times per optimal state. We repeat this algorithm for all 100 different optimal states. Finally, we average over the number of Monte Carlo steps using Eq. (3).

$$K_n = \frac{\sum_{j=1}^{u=100} \sum_{i=1}^{w=100} k_i}{u \times w}, \qquad (3)$$

where $k_i$ is the number of Monte Carlo steps before an optimal state is abandoned. Figure 2. Displays $K_n$ for systems with different efficiencies. Each point in Figure 2. belongs to a different random initial subgraph with the same range of efficiency in their optimal states and efficiency values are averaged over all these initial state and $u$=100 corresponding optimal states.

## 3.1 Resistance

We start our interpretation of the results in Figure 2. by the observation that for a given noise (perturbation) higher values of $K_n$ imply higher stability and vice versa. Networks with higher efficiencies need more steps to deviate from its optimal state and exhibit a higher level of resistance against external perturbation.

The effect of noise on our experiment is understood by comparing Figure 2. (*a* and *b*). For a given system with averaged efficiency of 0.6, $K_n$ is around 200 for small noise of $T = 0.01$ whereas for larger noise of $T = 0.15$ only around $K_n = 4$ to 5 steps are enough for the system to lose stability. More importantly, all the plots in Figure 2. demonstrate a linear relation between efficiency and $K_n$. This means that efficiency and stability have a linearly dependence. In other words, highly efficient systems are more resistant to external perturbations.

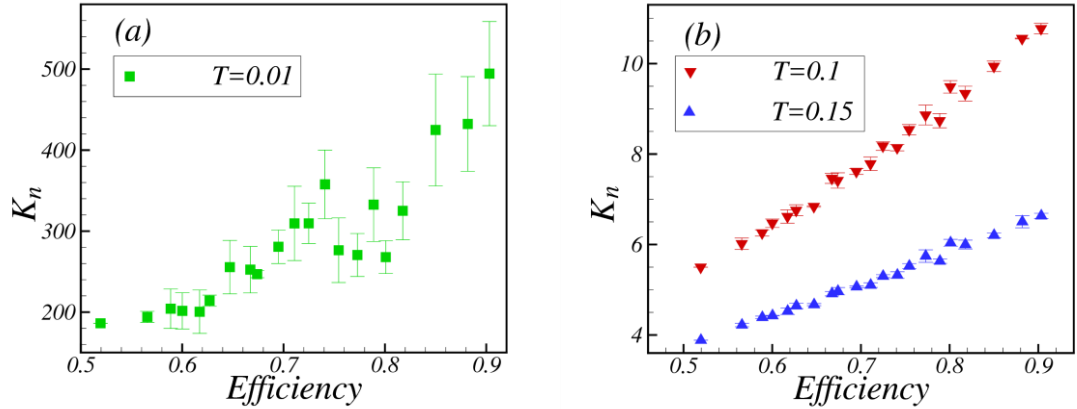

**Figure 2.** $K_n$ **(the number of steps that noise is turned on until the first change in efficiency occurs) versus efficiency** Data correspond to local optimal states in three different temperatures and show a linear dependency of stability as a function of efficiency. Highly efficient networks are more stable against perturbations: (*a*) $T = 0.01$ for small values of noise (temperature) the $K_n$ values (the number of steps needed to "kick out" the system from its local optimal state) are much larger than (*b*) for increased magnitude of the perturbations, i.e., $T = 0.1$ & $0.15$, values of noise.

Next, we study the reaction of a network to external perturbation. In particular, we would like to measure the change in the efficiency upon perturbation as a function of the efficiency itself.
As in Ref. [23], it is assumed that the system is located in one of its fixed (here: metastable) points, and we are interested in the question of how a single function (here the efficiency and the $GRC$) behaves if external perturbations (here increasing the temperature) are added. We consider further perturbations in the form of targeted node removal. Among others, we would like to measure the change in efficiency after perturbation is applied as a function of system efficiency.

In Figure 3. the average change in the efficiency $\langle \Delta E \rangle$ is shown as a function of efficiency for two values of noise (*T*). There is an approximately linear dependence between $\langle \Delta E \rangle$ and the efficiency. For systems with higher efficiency the absolute value of $\langle \Delta E \rangle$ (reaction) of the system is larger in response to a perturbation. Since we already established that higher efficiency translates to higher stability we can conclude based on our findings so far the following: Systems with higher stability are less susceptible to external perturbation but once the perturbation is large enough, they undergo a more pronounced change.

Below we show that for large values of the noise (perturbation), the reaction of the system ($\langle \Delta E \rangle$ and $\langle \Delta GRC \rangle$) is very similar to the randomly relocation of an edge in a network that operates at its optimal state. Random relocation here implies removal of an edge and adding it between two disconnected nodes in a random way. Figure 4-a shows the change in the efficiency for a large noise $T=5$ (red curve) and for random relocation of an edge (green) as a function of efficiency of the corresponding optimal states. Similarly, Figure 4-b, shows $\langle \Delta GRC \rangle$ as a function of $GRC$ for two perturbation approaches shown by the red and green data points.

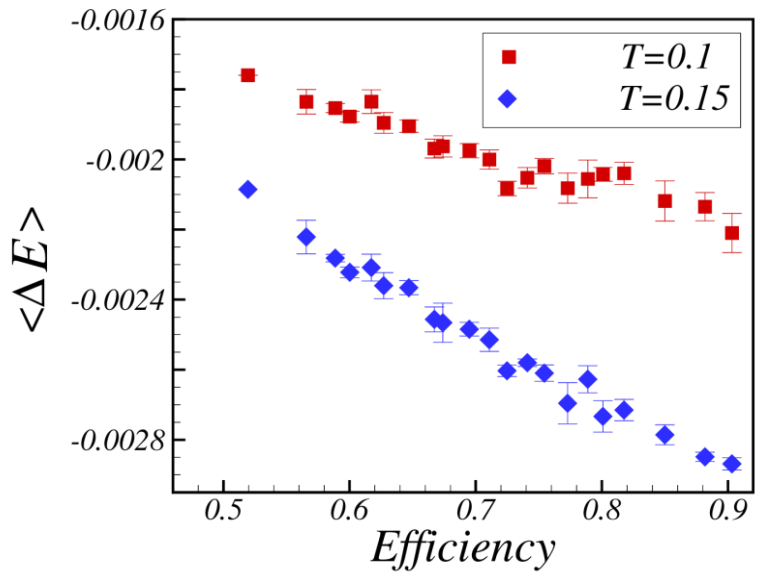

**Figure 3. Average of efficiency difference $\langle \Delta E \rangle$ between optimal and unstable states versus efficiency in two different values of noise (temperature).** There is a larger decrease in the efficiency for optimal networks with larger efficiency.

The fact that the behaviour of the efficiencies is similar when large noise or random relocation is applied as a perturbation is consistent with our expectations of what should happen for the case of large noises, since in the latter case a new edge is chosen almost randomly due to the larger value of the Boltzmann factor.

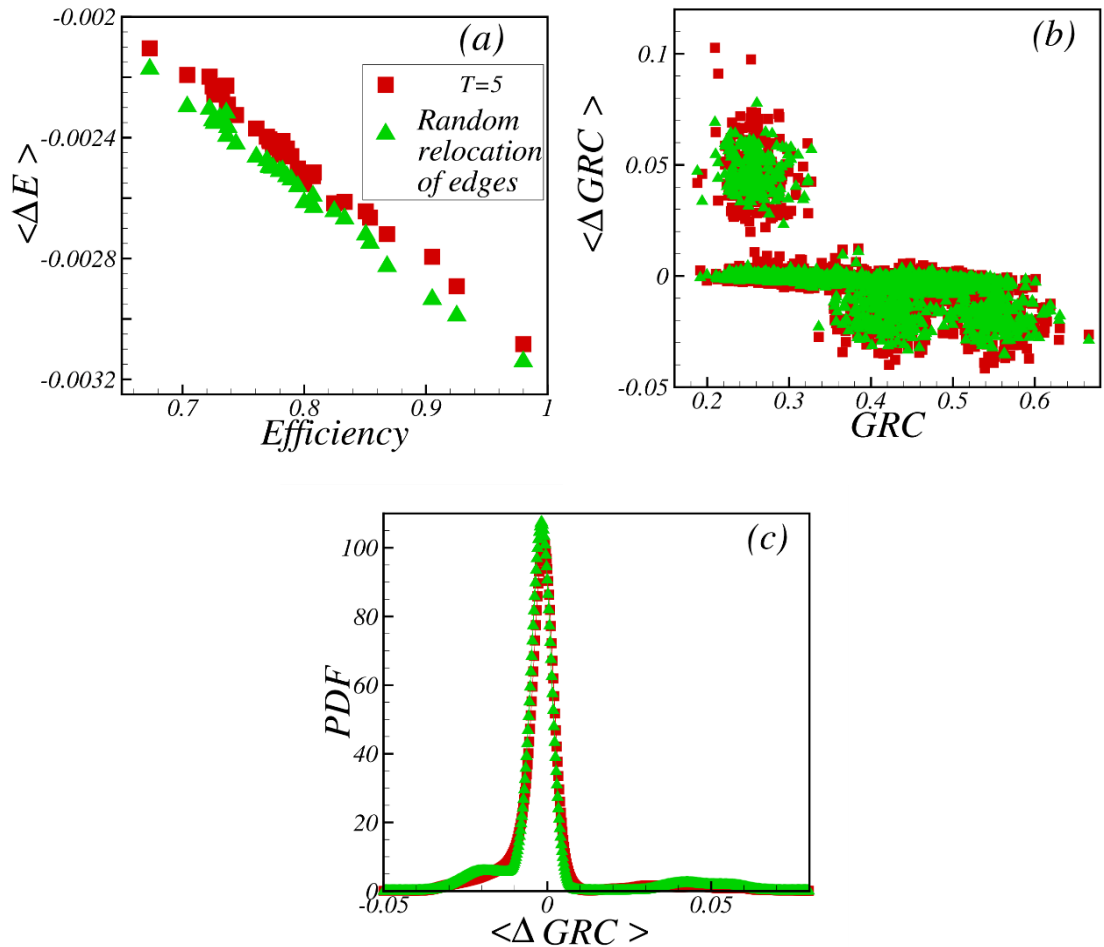

**Figure 4. Comparison of the effect of random relocation of edges in optimal states and high value of noise (*T=5*).** The reaction of the optimal networks is the same in both approaches. (*a*) ⟨Δ$E$⟩ versus $E$, (*b*) ⟨Δ$GRC$⟩ versus $GRC$ (*c*) Probability density function of ⟨Δ$GRC$⟩ centered on 0, represents the high resistance of hierarchical structure against eternal noise.

Figure 4-b displays two regions: In less or non-hierarchical networks with small $GRC$, the effect of perturbation makes the graph deviate to higher $GRC$ states thus ⟨Δ$GRC$⟩ is positive. For states with large $GRC$ the ⟨Δ$GRC$⟩ values are negative indicating a jump to the less hierarchical state (i.e., perturbations are likely to decrease the otherwise high level of hierarchy corresponding to high level of efficiency). Figure 4-c demonstrates the probability density function of ⟨Δ$GRC$⟩ with a high peak around ⟨Δ$GRC$⟩ = 0 and a side shift to the negative values which indicates that hierarchical optimal states have a higher resistance against external perturbation, preserve their structure and in the case of noise effect they lose their stability by jumping to the less hierarchical states. Figure 5. depicts the probability density function of ⟨Δ$E$⟩ and ⟨Δ$GRC$⟩ in four different temperatures.

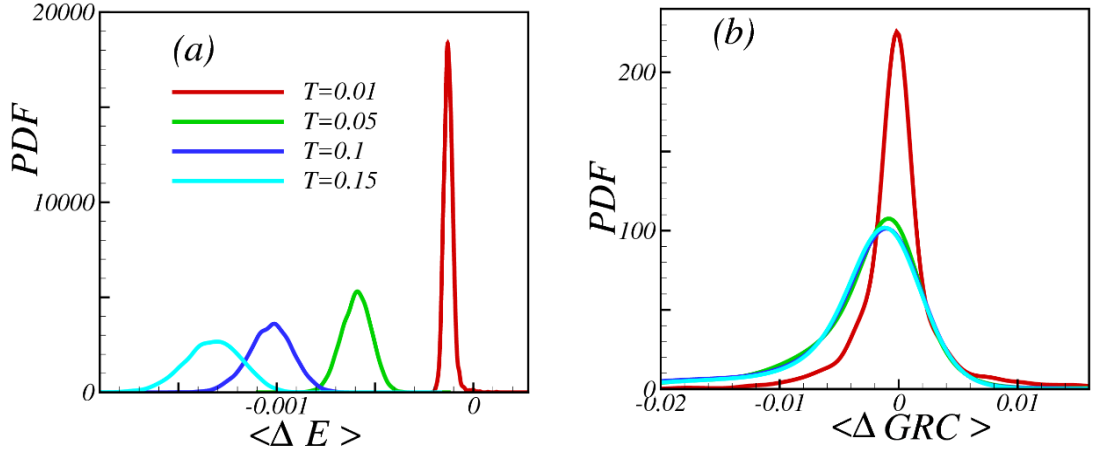

**Figure 5.** (*a*) Probability density function of ⟨Δ$E$⟩, optimal networks deviate to unstable states with less efficiencies and by increasing noise, the absolute value of efficiencies increase (*b*) Probability density function of ⟨Δ$GRC$⟩ in different temperature. In all temperature, the peaks are around zero demonstrates stability of networks structure in optimal states.

### 3.2 Resilience

The network's ability to retain its basic functionality after external perturbation and returning to its optimal state is studied in this section. To model this, we start with an optimal state and turn on the noise changing $T$ from nearly zero to $T$=0.1. After 32 steps, noise is switched off and the system is allowed to recover from its unstable state and converge to a local optimal state again. Figure 6. demonstrates efficiency and $GRC$ values in the whole process from the time that noise is turned on (step=0) and off (step=32) until the network saturates to its local optimal state. According to Figure 6-a when noise is turned on, the system experiences a sudden decrease in both the efficiency and $GRC$. Larger values of noise make stronger disturbance to the system. After the noise is switched off, the efficiency and $GRC$ increase again and converge to a higher value corresponding to one of the local optimal states.

We then calculate efficiency and $GRC$ differences between the new state and the initial local optimal state. Figure 7-a demonstrates variation of ⟨Δ$E$⟩ versus efficiency for three different initial complete graph. The linear trends with the negative slope in Figure 7-a show that as the efficiency increases the difference between two local optimal states are decreased. For larger efficiencies the network returns to the same initial optimal states ⟨Δ$E$⟩ = 0 after turning off the noise. This confirms the high resilience of highly efficient networks. For less efficient states, the difference of two optimal states are larger. Figure 7-b shows probability density function of <Δ$E$> with positive skew and a high peak close to zero.

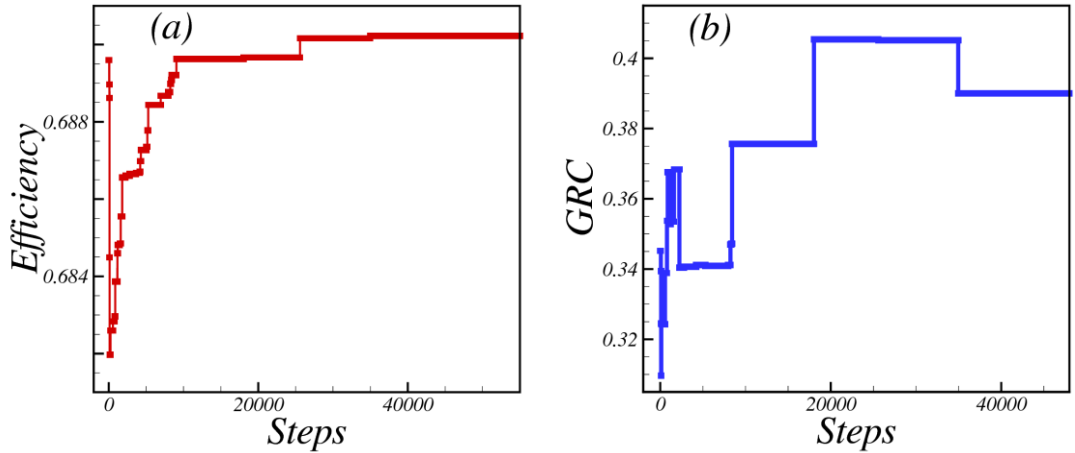

**Figure 6.** (*a*) **Variation of Efficiency during the effect of external perturbation.** First the perturbation $T = 0.1$ is applied and then the noise is switched off until the efficiency saturates. (*b*) Change of $GRC$ during the perturbation and after turning off the noise. Noise is turned off after 32 steps and the system saturates to other optimal state with higher efficiency.

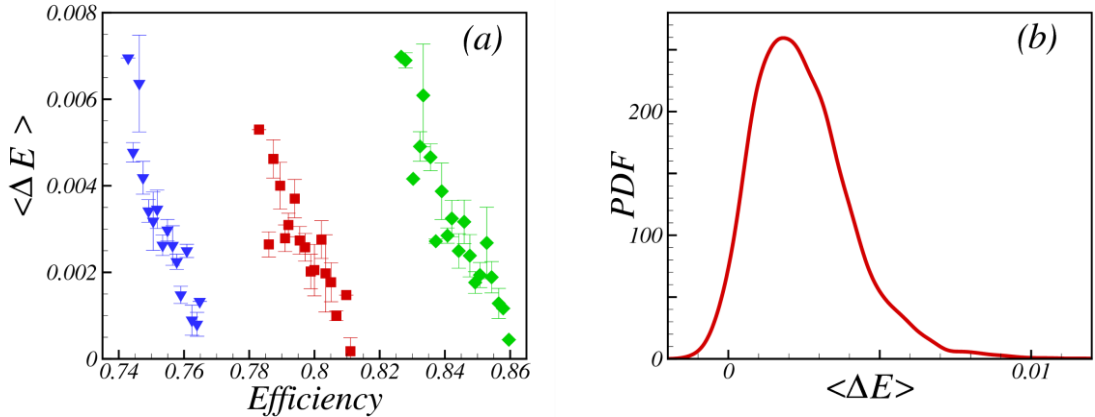

**Figure 7.** (a) **Average efficiency difference between two local optimal states $\langle \Delta E \rangle$ before turning on the noise and after its turn off versus efficiency ($E$) for three different initial conditions.** After the noise is switched off the network saturates to higher efficient state, $\langle \Delta E \rangle$ is positive. For each initial state higher efficient states exhibit higher resilience: $\langle \Delta E \rangle$ converges to 0 and the network converges to the same initial optimal graph. (b) Probability density function of $\langle \Delta E \rangle$ is skewed with a peak close to zero, demonstrating the resilience of some of the efficient states and positive values, i.e., switching to a more efficient state.

### 3.3 Resistance against node removal

In this section, we study the resistance of optimal networks against targeted node removal. Consider, e.g., a robotic network with $N$ robots (nodes) and edges representing existing links among robots, as proposed in Ref. [9]. Such links are intermittent during the execution of a real world mission, since the reach, between two nodes, varies in time given current communications technologies. During the execution of the mission, some nodes might have their battery drained or even in some cases, some robots might be destroyed or experiment all sort of failures. Also, the communication among the robots is not necessarily symmetric. Thus, during a mission carried out by a network of robots of varying function the underlying structure of the signals sent within the system can indeed be interpreted as a hierarchical directed network. A possible further interpretation is in which the nodes represent tasks to be completed by a group of robots and edges represent dependency among tasks. Another network example, in the realm of society, is a military organization, i.e. an army. Clearly, it is possible to describe an army as a directed hierarchical network. Therefore, what could happen if a general or a high-ranked military person is lost during combat? Alternatively, what happens if low-ranked soldiers are lost during a combat? We use our efficiency function to evaluate the effect of attacks on the networks we consider and suggest that the basic features we observe are likely to be applicable to other hierarchical systems as well.

In order to observe the effect on stability when nodes are lost or removed, we performed the following numerical experiments. We define an attack as the removal of $Q$ nodes, being $Q = [1, 2, 3, 4, ..., N]$, i.e. attack $Q$ consists of removing $Q$ nodes at once. We start with one local optimal state and perform $Q$ attacks and after each attack efficiency and $GRC$ of the network are measured. Given the hierarchical structure of the networks studied here, a finite set of different $C_{R(i)}$ (Local Reaching Centrality) exists for the nodes in the network i.e. $C_{R(i)} = [C_R^1, C_R^2, ..., C_R^x]$ with $N_1$ nodes having $C_R^1$, $N_2$ nodes having $C_R^2$, etc. s. We investigate the stability of the networks in their local optimal states by removing nodes (attacks) using two approaches. First, nodes with higher $C_{R(i)}$ and their corresponding links are removed one by one, and this process is continued to lower $C_{R(i)}$ values, while the efficiency and $GRC$ of the networks are measured after each attack. In the second approach, node removal starts from those with lower $C_{R(i)}$, and it continues to the higher ones. For a specific attack, each of the removed nodes possess a given $C_R(i)$. From hereafter instead of $C_R(i)$, the symbol $LRC$ is used. Thus, the $LRC$ -s are $LRC = [LRC_1, LRC_2, ..., LRC_x]$ with $N_1$ nodes having $LRC_1$, $N_2$ nodes having $LRC_2$, etc., $Q = [1,2,3, ..., 128]$ nodes were removed from the network in two approaches, from highest towards lowest $LRC$ and vice versa.

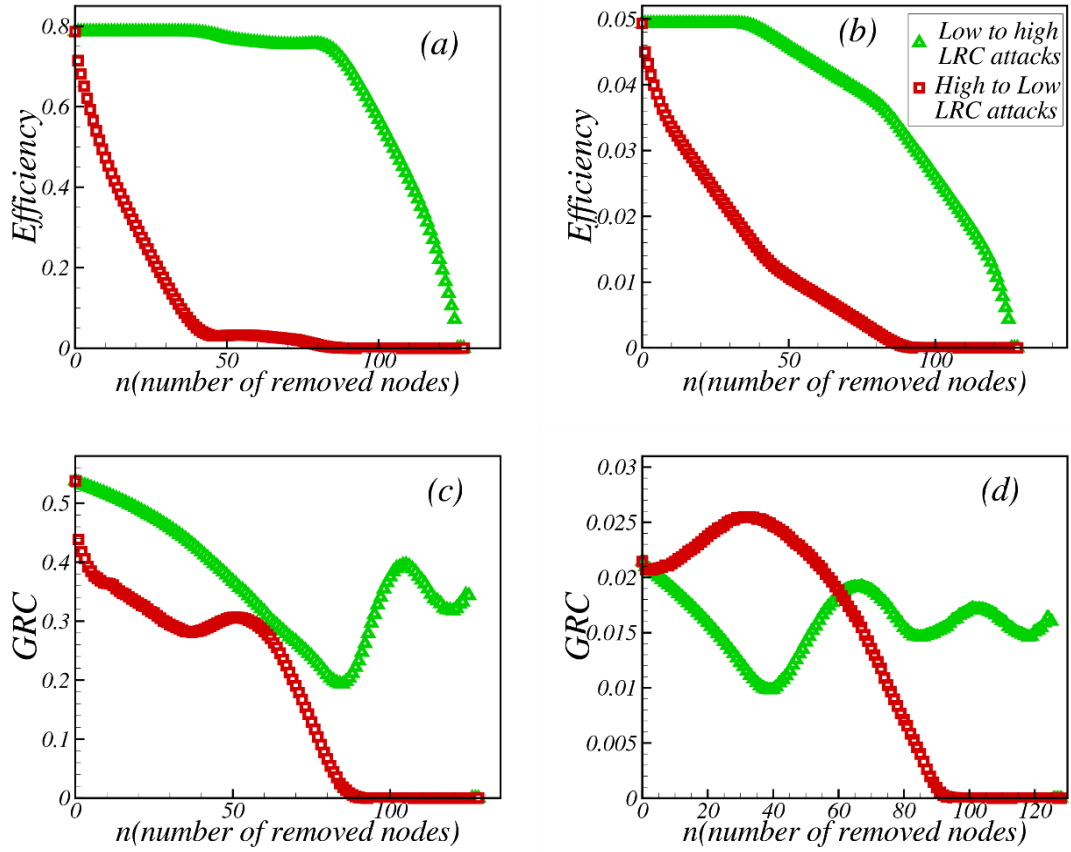

**Figure 8. Efficiency and $GRC$ after the attacks for networks with high (a, c) and low (b, d) $GRC$.** High $GRC$ corresponds to $GRC$-s in the interval [0.5, 1.0] and low $GRC$ corresponds to $GRC$-s in the interval [0.0, 0.5]. For $Q = n$ nodes, the *y*-axis represents optimal values for efficiency. The fluctuation-like behaviour of the plots (especially in c and d) is due to the finite (relatively small) size of the networks having only a discrete set of $LRC$ -s.

With the objective of analysing stability for networks with low $GRC$ (lower than 0.5) and high $GRC$ (higher than 0.5), under perturbations by node removal, we performed $Q$ attacks, with $Q = [1, 2, ... , 128]$ nodes, for 32 networks (18 with high $GRC$ and 14 with low $GRC$), each with 16 local optimal states. The nodes were ordered by two different approaches: *1. highest to lowest LRC* (red curves) and 2. *lowest to highest LRC* (green curves). This way, each attack $Q$, consisting of removing $Q$ nodes, was done according to these two methods. Figure 8-a shows that when nodes, in networks with high $GRC$, are removed with approach one (highest to lowest $LRC$), efficiency decreases faster than when nodes are removed with approach two. This is expected according to the model presented in Ref. [10], where nodes, with high ability and correspondingly high $LRC$, contribute to the efficiency to a greater extent and are at the top layer of the corresponding hierarchical network than nodes with low ability.

Similar behaviour appears for networks with low $GRC$ as depicted by Figure 8-b. However, in approach two (lowest to highest $LRC$), more than 80, out of 128 nodes removed causes the efficiency to drop significantly for high $GRC$ networks, whilst more than 40 nodes removed causes efficiency to drop significantly for low $GRC$ networks. It is clear from Figure 8 that the two kind of attacks lead to rather different outcomes for a large range of the removed nodes.

These results suggest that networks with high $GRC$ (more hierarchical) are very stable and efficient, even when losing large quantity of nodes, in comparison with low $GRC$ networks. Figure 8-d shows fluctuating behaviour for $GRC$, with attacks according to approach two (lowest to highest $LRC$). These fluctuations show that networks with low $GRC$ are not stable against external perturbations, such as node removal. Again, there is not relevant $GRC$ increase in approach one.

While the experiments, performed in this work, are over networks whose efficiency corresponds to that in the model presented in Ref. [10], $GRC$ is calculated according to the general model presented in Ref. [24], which suggest that these results are applicable to general hierarchical directed networks.

**Acknowledgements:** This research was supported in part by the grants USAF Grant No: FA9550-17-1-0037 and EU FP7 RED-Alert No: 740688

**References:**

[1] Thebault E and Fontaine C 2012 Stability of ecological communities and the architecture of mutualistic and trophic networks *Science* **329** 5993

[2] Neutel A M, Heesterbeek J A P and De Ruiter P C 2001 Stability in real food webs: Weak links in long loops *Science* **296** 1120

[3] Lorite A G, Guimera R and Pardo M S 2016 Long-term evolution of email networks: Statistical regularities, predictability and stability of social behaviours *PloS one* **11** e0146113

[4] Skyrms, B and Pemnatle R 2000 A dynamic model of social network formation *PNAS* **97** 16

[5] Kelly F P, Maulloo A K and Tan D K H 1998 Rate control for communication networks: Shadow prices, proportional fairness and stability *Journal of the operational research society* **49** 3

[6] Nisan N, Roughgarden T, Tardos E and Vazirani V V 2007 Algorithmic game theory *Cambridge University Press*

[7] Dutta B and Jackson M O 2000 The efficiency and stability of directed communication networks *Review of Economic Design* **5** 3

[8] Jackson M O 2008 Social and economics networks *Princeton University Press*

[9] Camargo-Forero L, Royo P and Prats X 2017 On-board high performance computing for multi-robot aerial systems in intech (Ed.) *Aerial Robots* In press
ISBN: 978-953-51-5357-3

[10] Zamani M and Vicsek T 2017 Glassy nature of hierarchical organizations *Scientific Reports* **7** 1382

[11] Mezard M, Parisi G and Virasoro M A 1987 Spin glass theory and beyond *World Scientific*

[12] Mezard M. and Montanari A 2009 Information, Physics, and computation *Oxford University Press*

[13] Stein D L and Newman M C 2013 Spin Glasses and Complexity *Princeton University Press*

[14] Lyapunov A M 1992 The general problem of the stability of motion *Taylor and Francis*

[15] Callaway D S, Newman M E J and Strogatz S H, D. J. Watts 2000 Network robustness and fragility: Percolation on random graphs *Phys.Rev.Lett.* **85** 25

[16] Cohen R, Erez K, ben-Avarham D and Halvin S 2000 Resilience of the internet to random breakdowns *Phys. Rev. Lett.* **85** 4626

[17] Matisziw T C, Grubesic T H and Guo J 2012 Robustness elasticity in complex networks *PloS one* **7** e39788

[18] Albert R, Jeong H and Barabasi A L 2000 Error and attack tolerance of complex networks *Nature* **406** 378

[19] Ghoshal G and Barabasi A L 2011 Ranking stability and super stable nodes in complex networks *Nature communications* **2** 394
[20] Jackson M O and Wolinsky A 1996 A strategic model of social and economic networks *Journal of economics theory* **71** 1
[21] M O Jackson, A Watts 2002 The evolution of social and economic networks *Journal of economic theory* **106** 2
[22] Konig M D, Battiston S, Napoletano M and Schweitzer F 2012 The efficiency and stability of R&D networks *Games and economic behaviour* **75** 2
[23] Gao J, Barzel B and Barabasi A L 2016 Universal resilience patterns in complex networks *Nature* **530** 307
[24] Mones E, Vicsek L and Vicsek T 2012 Hierarchy measure for complex networks *PloS one* **7** e33799